# Outlining the Design Space of Explainable Intelligent Systems for Medical Diagnosis

**Yao Xie**
UCLA, ECE
Los Angeles, California
yaoxie@g.ucla.edu

**Xiang 'Anthony' Chen**
UCLA, ECE
Los Angeles, California
xac@ucla.edu

**Ge Gao**
University of Maryland, iSchool
College Park, Maryland
gegao@umd.edu

## ABSTRACT
The adoption of intelligent systems creates opportunities as well as challenges for medical work. On the positive side, intelligent systems have the potential to compute complex data from patients and generate automated diagnosis recommendations for doctors. However, medical professionals often perceive such systems as "black boxes" and, therefore, feel concerned about relying on system-generated results to make decisions. In this paper, we contribute to the ongoing discussion of explainable artificial intelligence (XAI) by exploring the concept of explanation from a human-centered perspective. We hypothesize that medical professionals would perceive a system as explainable if the system was designed to think and act like doctors. We report a preliminary interview study that collected six medical professionals' reflection of how they interact with data for diagnosis and treatment purposes. Our data reveals when and how doctors prioritize among various types of data as a central part of their diagnosis process. Based on these findings, we outline future directions regarding the design of XAI systems in the medical context.

## ACM Classification Keywords
H.5.m. Information interfaces and presentation (e.g., HCI): Miscellaneous.

## INTRODUCTION
Intelligent systems, the computational agent that employs algorithms to process and make sense of data, are becoming increasingly ubiquitous in modern workplaces [1]. Despite the promise of assisting human decision making through a data-driven approach, non-computing professionals often find it challenging to understand how the system transforms their initial input into a final decision and why.

In the medical field, systems such as the CheXNet [40] have been developed to interpret a patient's chest X-ray scan using deep learning. While the system can perform faster than human doctors with impressive accuracy, it offers little clue to indicate what happens within the "black box". Human doctors holding medical responsibility can hardly trust the system's results without understanding its underlying decision-making process [21].

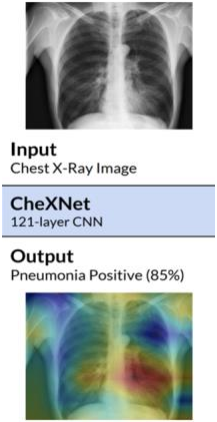

**Input**
Chest X-Ray Image

**CheXNet**
121-layer CNN

**Output**
Pneumonia Positive (85%)

**Figure 1. The input and output image of CheXNet [40]**

To help non-computing professional better comprehend results generated by intelligent systems, a growing body of research has been conducted with the goal of building explainable AI (XAI). It provides various *system-centric* solutions, such as developing accountable and transparent algorithms [11,43], visualizing obscure features [12,49], and employing theories from cognitive psychology to explore effective explanations [28,29,32]. The current limitation of these approaches is that there is a lack of empirical evidence to support the understanding by domain professionals [24].

In this project, we tackle the challenge of XAI from a *user-centric* perspective. We identify medical domain as the focus of our research given the proliferation of AI-powered diagnosis systems in recent years. We hypothesize that human doctors will find a system more explainable when the system 'speaks the language' of a doctor and 'thinks like' a doctor,

The remainders of this paper present our first step to the design of an explainable AI system by taking the perspective of medical professionals. We firstly review prior research on XAI, intelligent system in the medical field, and mental model of medical professionals, respectively. After that, we report a preliminary interview study with six doctors that tells how medical professionals interact with data for diagnosis and treatment purposes in their daily work practice. Based on findings from the interview, we discuss how interaction designers can incorporate human doctors' data processing model into medical intelligent systems and make such systems more explainable for the users.

## BACKGROUND & RELATED WORK
In this section, we first lay out a background review on XAI research, and then zoom into an HCI-oriented approach towards XAI. Since our focused field is in medicine, we further discuss prior work in medical AI, and specifically related to our interest—literature on the reasoning process of medical professionals.

**Explainable Artificial Intelligence (XAI) Systems**

Explainable artificial intelligence (XAI) raised a lot of concerns in recent years [20]. Since 1970, researchers have focused on the explanations of expert intelligent systems [31,46]. Recently, the need for explainable artificial intelligent is called for again because of the development of machine learning and artificial intelligence. Especially, algorithms like deep learning are intrinsically difficult to be understood and it brings the need for better explainable systems.

A lot of work of interpretable machine learning has been done to explain the inner principles of the machine learning models with mathematical and algorithmic solutions [4]. The main methods of interpretable machine learning are explanations of the complex algorithm like deep learning, causal inference, Bayesian rules, and visual analytics. Algorithm accountability means that the algorithm should explain the decisions. For example, "right to explanation" law in EU [18]. Planning oversight, retrospective analyses, and continuous review are needed to make the algorithm accountable[19]. However, there are still many challenges in XAI. Lipton [27] proposed a taxonomy of the reasons for interpretability and also the ways to interpret but there is still no consensus about the definition of interpretability. Some researchers studied the evaluation of whether a system is interpretable and evaluation methods are proposed [13]. Attempts have also been made to map the intelligibility, interpretable algorithms and explainability with the related work. In social science, researchers also study how people define, select, generate, evaluate and express an explanation [33].

**Intelligibility and Explainable Systems Research in HCI**

In HCI, researchers are focusing on user's interaction with the intelligent systems and explanation is one important topic. HCI researchers focus more on the interaction between the artificial intelligent system and users and they have done a lot of work from this aspect. Artificial intelligent systems have been criticized that their rigid concepts are incompatible with human behavior styles [45]. Explainable artificial intelligence in HCI contains topics including context awareness, cognitive psychology, and software learnability [44]. Context awareness is used to recognize user reactions and activities. In the early 2000s, context awareness has raised a lot of concern with the development of mobile devices and sensors [9,42]. People should understand what is sensed and what reaction is taken under a specific situation. For a context-aware system, it should let users know "what they know, how they know it and what they are going to do next" [3]. The needs for simplistic representations of the context in explainable AI is called to let users be aware of what is obtained and which action will be done by systems [14]. Cognitive psychology is more about explanation theory. Lombrozo studied cognitive explanations [28] and found that it is strongly connected with causality reasoning. Also, XAI not only focuses on human cognitive psychology but also the understanding of social context [33]. Software Learnability is an important part of usability. It focuses on how to use complex software applications with the help of demonstrations or in-context videos [19] and it evaluates the easiness of using a system.

Systems need to provide users with not only results but also the account of their behaviors [3]. Furthermore, research about a tailored interface that provides the visual or textual explanation for context-aware rules has been done [10]. Researchers also studied the design strategies of interaction and how to help users predict system behavior through feedforward [2,3,47,48]. How users understand and control the machine learning programs is also a relevant trend, which also works towards the debuggable and intelligible machine learning [50]. Understandability and predictability are very important in artificial intelligence applications such as autonomous vehicles [37]. Besides the algorithmic accountability, transparency, and fairness, data visualization is also a stream from the computational perspective of HCI, which seems to be isolated from what machine learning researchers do [7].

**Intelligent Systems in Medical Fields**

In medical fields, artificial intelligent systems also have a broad prospect. With the growth of availability of medical data and the data processing techniques, artificial intelligent systems are possible to be applied in the healthcare domain. They are able to dig out useful information from a large amount of data which is difficult to be processed by doctors and thus, assist the medical decision making [16,35]. In the medical field, it has three major applications: early detection, diagnosis, and treatment plan. They can also help with the diagnostic process of diseases such as cardiology, cancer, and neurology [23]. The research in medical artificial intelligence mainly focuses on pathology and radiology. For example, systems are able to identify the radiographs and recognize patterns for radiologist and pathologist and work as an information specialist during the diagnostic process [22]. Besides the image analysis applications in radiology and pathology, artificial intelligent systems are also applied to read the medical scientific literature and integrate electronic medical records. In addition, they may optimize and predict the treatment of chronic disease [32].

However, comparing to the booming industry, the actual usage of the autodiagnostic system in hospitals is relatively low. A study has been made to know doctors' acceptance and the adopt intention of these systems [15]. Another research proposed the methods of evaluating the clinical performance and effect of the artificial intelligent systems in medical diagnosis and one of the methods mentioned the explanations [39].

The explanation capabilities of artificial intelligence systems using knowledge bases are firstly added for the applications in medical decision making and computer-aided diagnosis in 1983 [46]. After that, a diagnostic reasoning theory is used to find the components of systems that lead to and explains the discrepancy between the expected result and observed behaviors [41] and it has a variety of settings such as medical auto-diagnostic systems. Further, how doctors make decisions under uncertain and information overloaded cases raises a lot of concerns. An argument-based interaction that is flexible and easily understood by human users is proposed to help doctors make decisions based on this question [17]. It is also proved that a fuller explanation has a positive effect on users' trust of such systems and also helps to solve reliance issues. Better explanations can let users better understand the reasoning chain, thus enhancing the system's confidence and help doctors provide better diagnoses [5]. An interactive visual analytics system is also designed to help support interactive dependence diagnostics by feature representation and visualization [24].

**Medical Reasoning, Decision Making & Mental Models**
Cosby summarized tow models of clinical reasoning: analytical and intuitive [8]. The analytical approach is based on the hypothetic-deductive model that is common in scientific research and discovery, whereas the intuitive approach is akin to recognizing common patterns from a patient's symptoms rather than deliberately going through a methodological decision-making process. Doctors often choose one of these models based on how experienced they are and how complicated a case is.

Also, due to the uniqueness of the medical field, medical reasoning and decision-making mean more than what they mean in other fields. From the doctors' perspective, explanation of the decision making process is not only how the results come out, but also the cost of medical decisions such as the responsibility and risk [6]. In different scenarios, the requirement of explanations also varies. In addition, the decision-making process in the medical field can be regarded as a combination of basic medical knowledge such as pathology, the experience gained by previous patients in similar conditions and the cognition of the patient's demographic information. It is a lot more complex than regular decision-making process and mental model which can be reached by splitting different features with "yes" or "no" [8].

Broadly, the term 'mental model' is a concept derived from cognitive psychology. It is the explanation of people's thought process about how things work [38]. The mental model can also be regarded as an internal representation of the external factors and it is important in cognition, decision making, and reasoning [36]. The internal conceptualizations including users' beliefs and understanding about the system behavior will guide their interaction with the systems [38]. Also, during the interaction, the mental models will develop individually according to different users. In general, most mental models are simpler than the actual systems and it is sufficient to allow users to understand the system behavior [34]. However, when it comes to the complex cases, for example, medical diagnosis, if mental models cannot reflect the actual complexity of these artificial intelligent systems, users might feel difficult to understand, explain or predict the system behavior [38]. In order to make users better understand and explain how the system works, the system should be transparent and show the mental model similar to human's mental models [25,26]. Otherwise, users are likely to build flawed mental models when interacting with such systems and be confused about the process of decision making [38]. For systems with improved mental models, user's satisfaction perceived control, and the overall trust of the system will all be enhanced, which will also facilitate understanding [25].

**INTERVIEW**
Even though a lot of researches have been done to explain the intelligent systems. They seldom look into specific domains and incorporate empirical knowledge when explaining. We try to understand this problem from the doctors' perspective and that's why we seek to investigate the following research question:

**RQ:** How do medical professionals interact with patients' data for diagnosis and/or treatment purposes?

**Overview**
We conducted an interview study to explore research questions presented above. Our current sample consists of six licensed medical professionals working in California, United States. Each interview lasts about 1 hour. During the research process, we iterated between collecting new data, generating codes, and revising/elaborating the existing coding book as suggested by the grounded theory [30]. Findings from these interviews offered insights revealing the relationship between medical professionals, data and intelligent systems from a human-centered perspective.

**Participants and Data Collection**
All interviewees joined this study by responding to an online participant call posted by the research team. We intentionally looked for participants who hold different domains of expertise within the medical field, so that the interview data can best capture both the commonalities and the differences between the thinking styles of various medical professionals. Table **1** summarizes the background information of each interviewee. For the anonymous purpose, we replaced their names by randomly assigned IDs.

Between September and November of 2018, the first author of this paper conducted semi-structured interviews with each participant. The interview protocol was initially developed through in-group brainstorming sessions among the authors of this paper. It then got revised based on two pilot interviews with senior M.D. students at UCLA. The final protocol consisted of questions revolving around four issues: 1) the interviewee's work and education experience in the medical field, 2) how s/he accesses to, processes and interprets medical related data during daily work practice, 3) challenges and solutions s/her ever experienced, if any, when working with medical data, and 4) experience and/or expectations of using computer-based systems to facilitate daily medical work. All interviews were conducted face-to-face in English and audio-taped for transcription.

| ID | Domain of Expertise/Specialty | Gender | # of Years in the Medical Field |
|---|---|---|---|
| P1 | Pathologist | Male | 22 |
| P2 | Orthopedist | Female | 17 |
| P3 | Neurologist | Male | 7 |
| P4 | Family physician | Male | 10 |
| P5 | General physician | Male | 5 |
| P6 | Cardiologist | Male | 18 |

**Table 1. Background information of our six interviewees, including their participant ID, the domain of expertise, gender, and number of years working/studying in the medical field**

**Analysis**

Three authors of this paper analyzed the interview data together following an inductive approach. There were 60 codes and 201 quotations generated from the initial open coding. They yield participants' self-reflection regarding the forms of data they interact with at daily medical work, the thinking process they go through when interacting with various data, the decisions they try to make based on data processing, and the types of work they have been delegating or hope to delegate to computer-based systems.

We reiteratively discussed and compared between codes as they were generated. During the discussion, *prioritization* emerged as a focal theme from the data. It indicates that a central task medical professional performs during diagnosis is to prioritize among various and sometimes conflicting information given by patients, other doctors, and computer-based systems. We went through further coding to identify connections between this focal theme and other emerged themes and categories. The following section presents our detailed findings. Words and phrases directly quoted from participants are written in italic.

**FINDINGS**

The process of generating a proper diagnosis and/or treatment plan is frequently described by our interviewees as being *context-dependent*, *data-intensive*, and *open to alternative possibilities*. In many cases, there lack one-to-one correspondences between signs, symptoms, and diseases. Medical professionals in the field, therefore, are often required to integrate various kinds of data and think outside the box. As it is pointed out by the following two participants:

*For medicine, it's usually the grey area that matters. Everything is hardly black and white, and that's why it is always difficult. ... People say that medicine is both a science and an art, because every disease is different, and every patient's representation will be different. Every doctor obviously has different steps in making the decision. [P6, Cardiologist]*

*Authorities, like the American Heart Association, will publish guidelines and flow charts that we can refer. It prevents physicians from making ridiculous mistakes. But for more complex diseases, the guideline cannot include all of them. It will depend on the doctor's experience or some innovations to accomplish the treatment. [P4, Family physician]*

In the rest of this section, we describe how medical professionals navigate around the complexity of their interaction with medical data. We identify three critical steps from interviewees' reflection, including detecting/reacting to borderline cases, generating prioritization matrices, and coordinating with computer-based systems. Across all these steps, medical professionals keep prioritizing and re-prioritizing among information collected at different stages of the diagnosis process.

**Borderline Cases: When Challenges Emerge**

All participants of our interview reported running into borderline cases as the moments when the processing and interpreting of medical data turn challenging. One representative situation of encountering borderline cases is when the symptoms are still in their early state:

*At the very early state [of cancer], it is difficult to tell if the cell is abnormal. The architecture is minimally disrupted. You may think it is abnormal, but you don't know whether it is malignance. We will show the cases to other colleges to get consents, or we have to say this case is inconclusive. [P1, Pathologist]*

In other situations of the borderline cases, medical professionals receive conflicting information that indicates different directions of the diagnosis:

*Many of us have run into cases when the MRI doesn't confirm [our diagnosis]. We think the problem is in the right brain, but the image shows nothing there. In that case, we may do the test again. We can also go back to the patient to ask them again, or we discuss with other doctors. [P3, Neurologist]*

To clear up the ambiguity as indicated by the two quotations above, doctors often need to cross-validate their initial evaluation of the patient by requesting further data. Our interview with the six medical professionals documented multiple types of such data, including but not limited to, the patient's demographic information, cardinal symptoms, results from further physical examinations and lab tests, historical data from reference groups, and evaluations given by other doctors.

Participants in our study yielded similar insights regarding how they deal with the rich yet complex medical data. Instead of following one hard rule of data processing, interviewees tend to weight/interpret each type of data differently based on their personalized prioritization matrices.

**Prioritization Matrices: Validity and Beyond**
We identified six parameters from participants' self-reflections that reveal how they perform data prioritization for diagnosis and/or treatment purpose. These parameters are labeled as below:

- *Theoretical validity.*    Robustness of connections between signs, symptoms, and diseases as proved by theories, medical textbooks, and guidelines;

- *Severity of consequence.*    Quality and quantity of potential consequences if the detected signs/symptoms get put aside at this moment; side-effects of a treatment; interactions between different treatments;

- *Time constraint.*    Timing; urgency; sequential order of taking care of different symptoms and diseases;

- *Domain of expertise.*    The extent to which the signs and symptoms connected to the doctor's specialty; the level of confidence in offering a candidate treatment;

- *Risk avoidance.*    Responsibility assigned to a specific doctor; power dynamics between junior vs. senior doctors;

- *Technical feasibility.*    The sensitivity of the measurement; reliability of the technique; the false positive/negative rate of symptom detection.

Participants often used *styles* to describe the detailed prioritization matrices held by different doctors. Similar to other dispositional attributes such as personality, the prioritization matrix of a medical professional is perceived as being self-aware and consistent across various diagnosis made by the same individual:

*The diagnosis depends on many factors –severity, possibility, consistency with the patient's history, and others. Some doctors will make the most severe issues on*

*the priority, others will make the most possible ones their priority. It depends on their perspective. It also depends on the time concern. For example, neurologists may have a longer period of diagnosis, but surgeons and ER doctors don't. [P2, Orthopedist]*

*Some doctors trust images [over other information], like MRI, to tell what's happening. About 80% of the time you would have good images. You are very confident about the diagnosis from the images. But I think most important information [to facilitate diagnosis] is what the patient tells you. It helps to track the patient's history. [P6, Cardiologist]*

Our interviewees sometimes referred to the personalized prioritization matrices (or *styles*) to explain the disagreement between diagnosis suggestions provided by different doctors (see Figure 2 for illustration).

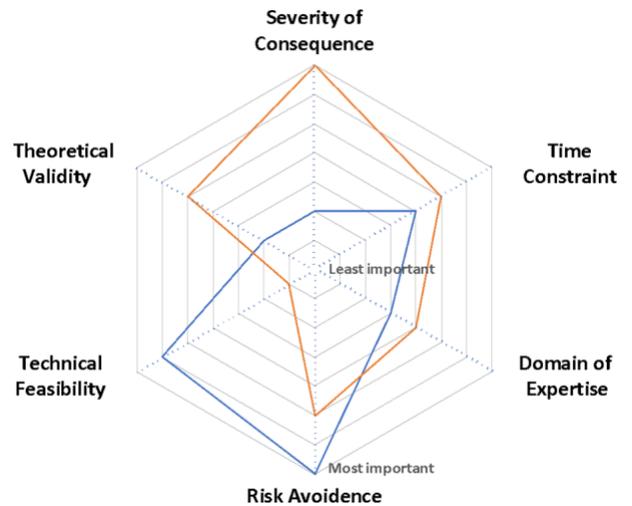

**Figure 2. A diagram that compares the prioritization matrices held by two different medical professionals (blue vs. orange line) when making diagnosis decisions. While one doctor uses severity as the primary parameter to weight various data during diagnosis /treatment, the other cares most about the calculation of risks and responsibilities.**

**Coordination Between Medical Professionals & Systems**
All medical professionals in our study reported that they have been using computer-based tools and systems to facilitate their daily work practice. Most participants, for instance, have greatly relied on cloud-based platforms to store and connect their local medical data with other databases [P1, P2, P4, P5, P6]. They also used various systems to generate automated calculation of chromosomes [P1], identify the degree of scoliosis [P2], check possible interactions between medications [P5], and etc. The primary function of such tools is to *"provide quantified information to doctors, but not [to give] answers in terms of high-level decisions* [P3]".

While participants were confident that the auto-quantified information given by systems is usually *trustworthy* and *helpful*, this optimism does not remain in their narratives of auto-diagnosis or treatment recommended by systems. The following quotation from P6 indicates a shared attitude as reflected across all the six interviews:

*There is a lot of advanced analysis involving machine learning, and some of them have entered the clinical realm. For example, you will have the nuclear images, and you will have the software telling you "it's abnormal here and there." It's as if you have a second reader next to you. I would love to have the system generating results, but ultimately, it's you that's deciding on the diagnosis. When there is a disagreement, me and everyone will be overwriting the machine-generated interpretation. [P6, Cardiologist]*

To step forward from quantifying information to directly assisting diagnosis and treatment, systems are expected to "*give an argument for why the data should be interpreted in that way [P5]*". The majority of our participants proposed the concept of *reference and comparison* as one approach to ground the systems' diagnostic reasoning with that of human doctors':

*Any machine has to give an evidence for the top reasons like in descending order for why in some matrices. It's like if I say something and you think differently, then we should be able to really compare the two. Otherwise, it doesn't matter if the machine's suggestion is right. I don't know what its thinking is and, ultimately, I take all the responsibility in this decision. [P1, Pathologist]*

*There are different ways [to help validate the systems' diagnosis recommendations]. One is showing me past examples in the database - will that support its conclusion? Another one is sources of data, something like research articles or convincing cases have been done. That's upper-level evidence. [P5, General physician]*

Interviewees further suggested that to build an ideal auto-diagnosis/treatment system, the algorithm should be able to *contextualize* its reference data with personalized information of a patient. Such contextualization work is what human doctors are good at based on their professional training and experience, but it is perceived to be the major obstacle for systems to overcome.

## IMPLICATION FOR DESIGN
Based on the findings of the preliminary interview, we outline design suggestions for explainable medical AI systems. Specifically, we envisage a system that can

- Allow a medical professional to prioritize different types and sources of data by directly manipulating a user interface akin to our proposed prioritization matrix (Figure 2);

- Support gradual engagement of medical AI systems into a medical professional's diagnosis process, spanning from low-level automated measurement tasks, to mid-level constraint-aware planning of medical tests, and to high-level suggestions of plausible diagnoses.

## ACKNOWLEDGMENTS
We thank Xiaohe Yang for assisting us to complete the interviews. We thank all the anonymous interviewees for their contributions to our study. We also thank Maie St. John, Peter Pellionisz and Jeff Liang for their valuable comments on earlier drafts of this paper.